\documentclass{article}

\usepackage{arxiv}
\usepackage[utf8]{inputenc} 
\usepackage[T1]{fontenc}    
\usepackage{hyperref}       
\usepackage{url}            
\usepackage{booktabs}       
\usepackage{amsfonts}       
\usepackage{nicefrac}       
\usepackage{microtype}      
\usepackage{graphicx}
\usepackage{doi}
\usepackage{subcaption}
\usepackage{tabularx}
\usepackage{rotating}
\usepackage{bm}  
\usepackage{amsmath}
\usepackage{cleveref}
\usepackage{natbib}
\usepackage{scalerel}

\usepackage{multirow}
\usepackage{enumitem}   

\title{Rule-Based Autocorrection of Piping and Instrumentation Diagrams (P\&IDs) on Graphs}

\author{ 
    \href{https://orcid.org/0000-0001-7494-9110}{\includegraphics[scale=0.06]{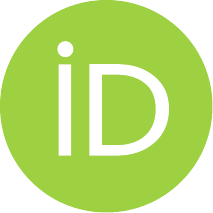}\hspace{1mm}Lukas Schulze Balhorn} \\
	Process Intelligence Research Group\\
	Department of Chemical Engineering\\
	Delft University of Technology\\
	\And
    \hspace{1mm}Niels Seijsener \\
	Fluor BV\\
	Amsterdam, The Netherlands\\
    \And
    \hspace{1mm}Kevin Dao \\
	Fluor BV\\
	Amsterdam, The Netherlands\\
    \And
    \hspace{1mm}Minji Kim \\
	Process Intelligence Research Group\\
	Department of Chemical Engineering\\
	Delft University of Technology\\
	\And
    \href{https://orcid.org/0000-0002-7837-055X}{\includegraphics[scale=0.06]{figures/orcid.pdf}\hspace{1mm}Dominik P. Goldstein} \\
	Process Intelligence Research Group\\
	Department of Chemical Engineering\\
	Delft University of Technology\\
	\And
    \hspace{1mm}Ge H. M. Driessen \\
	Fluor BV\\
	Amsterdam, The Netherlands\\
    \And
    \href{https://orcid.org/0000-0001-8885-6847}{\includegraphics[scale=0.06]{figures/orcid.pdf}\hspace{1mm}Artur M. Schweidtmann}\thanks{corresponding author} \\
	Process Intelligence Research Group\\
	Department of Chemical Engineering\\
	Delft University of Technology\\
    \texttt{A.Schweidtmann@tudelft.nl} \\
}

\renewcommand{\headeright}{}
\renewcommand{\undertitle}{}
\renewcommand{\shorttitle}{Autocorrection on P\&ID graphs}

\hypersetup{
pdftitle={Autocorrection on P\&ID graphs},
pdfauthor={L. Schulze Balhorn},
}

\begin{document}
\maketitle

\begin{abstract}
A piping and instrumentation diagram (P\&ID) is a central reference document in chemical process engineering. Currently, chemical engineers manually review P\&IDs through visual inspection to find and rectify errors. However, engineering projects can involve hundreds to thousands of P\&ID pages, creating a significant revision workload. This study proposes a rule-based method to support engineers with error detection and correction in P\&IDs. The method is based on a graph representation of P\&IDs, enabling automated error detection and correction, i.e., autocorrection, through rule graphs. We use our pyDEXPI Python package to generate P\&ID graphs from DEXPI-standard P\&IDs. In this study, we developed 33 rules based on chemical engineering knowledge and heuristics, with five selected rules demonstrated as examples. A case study on an illustrative P\&ID validates the reliability and effectiveness of the rule-based autocorrection method in revising P\&IDs.
\end{abstract}

\keywords{Autocorrection \and P\&ID graphs \and pyDEXPI}

\section{Introduction}
Undetected errors or suboptimal designs in Piping and Instrumentation Diagrams~(P\&IDs) can cause financial costs, hazardous situations, unnecessary emissions, and inefficient operation. These errors are currently captured in extensive manual revision processes leading to safe, operable, and maintainable facilities. However, engineering projects often involve hundreds to thousands of P\&ID pages, leading to a significant revision workload. With the advent of computer vision-based document digitization~\citep{Theisen2023_DigitizationChemicalProcess} and data exchange standards such as the Data Exchange in the Process Industry~(DEXPI)~\citep{Theissen2021_DexpiPidSpecification}, opportunities arise for algorithmic support of P\&ID development and revision~\citep{Schweidtmann2024_GenerativeArtificialIntelligence}.
Several studies have been conducted on error detection and correction in engineering diagrams. These studies can be grouped into rule-based approaches~\citep{Bayer2019_GraphBasedManipulation} and machine learning~(ML)-based approaches~\citep{Dzhusupova2023_UsingArtificialIntelligence,Oeing2023_GraphLearningMachine,MizanurRahman2021_GraphBasedObject,SchulzeBalhorn2024_AutocorrectionChemicalProcess}. Specifically, ML-based approaches function as follows: \citet{Dzhusupova2023_UsingArtificialIntelligence} detect erroneous designs using computer vision in P\&IDs. In contrast, \citet{Oeing2023_GraphLearningMachine} and \citet{MizanurRahman2021_GraphBasedObject} use node classification on P\&ID graphs to detect deviations from common P\&ID design. In addition, our previous work proposed a generative artificial intelligence approach that generates corrected P\&IDs using a string presentation~\citep{SchulzeBalhorn2024_AutocorrectionChemicalProcess}. The ML-based approaches offer the advantage of automatically learning to detect errors from existing P\&ID data. However, ML-based approaches require large-scale data for model training, currently lack explainability due to their black-box nature, and pose a risk of hallucination. Furthermore, ML-based methods do not provide a deterministic guarantee that known errors encountered during training will be detected during inference, posing a risk for safety-critical designs. Conversely, rule-based approaches, grounded in chemical engineering knowledge and heuristics, are beneficial for developing highly accurate error detection and correction methods without the need for large-scale P\&ID data. Rule-based approaches detect errors using hard-coded rules. These rules can be formulated based on general engineering logic or safety considerations. Moreover, the derived rules can be accompanied by an explanation. Whereas ML-based approaches still lack large-scale P\&ID data in practice, we expect that a rule-based approach is a good first step for a supporting tool for P\&ID revision because it builds trust with the engineer due to the high accuracy and explainability. 
Previous rule-based approaches represent P\&IDs as graphs and aim to detect errors by identifying abnormal graph patterns~\citep{Bayer2019_GraphBasedManipulation}. These approaches use graph-based rules to modify P\&ID graphs: Once an abnormal graph pattern is identified, specific modifications are applied to correct the P\&ID graph. While the previous work is an important contribution toward supporting engineers in P\&ID designs, we identify three limitations: (i) Previous methods are not linked to data exchange standards, making their integration into industrial P\&ID development difficult, (ii) the literature focuses on digitization errors but neglects engineering errors, and (iii) literature does not account for missing components in P\&IDs.
Building on the promising work by \citet{Bayer2019_GraphBasedManipulation}, we propose a rule-based autocorrection method for DEXPI P\&IDs to address the aforementioned limitations. The methodology can be summarized as follows: First, we use our pyDEXPI package~\citep{Goldstein2025_PydexpiPythonFramework} to convert a smart P\&ID into a graph. Second, rules are formulated based on chemical engineering knowledge and heuristics. These rules are implemented as graphs that describe manipulations to the P\&ID graph. Lastly, applying these rule graphs to the P\&ID graph completes the autocorrection process.

\section{Rule-based autocorrection methodology}

\begin{figure}
    \centering
    \includegraphics[width=\linewidth]{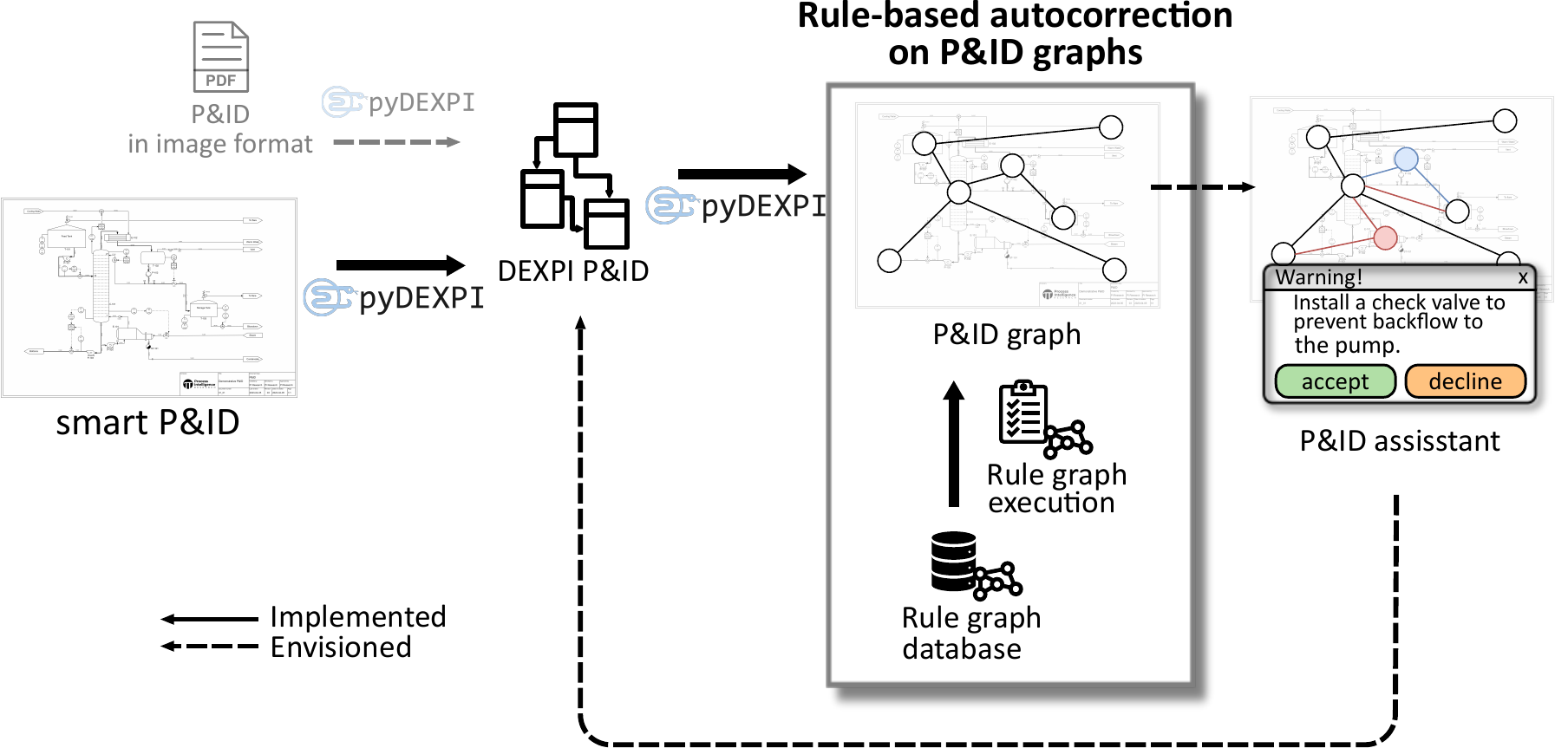}
    \caption{Rule-based autocorrection on graphs. We first translate a machine-readable smart P\&ID to a P\&ID graph. Then, we apply the engineering rules in the form of graph manipulations to the P\&ID graph. In future applications, the autocorrection method may be implemented in a P\&ID assistant with the human-in-the-loop. Changes can then automatically be updated in the DEXPI P\&ID. We envision applying the rule-based autocorrection also on P\&IDs in image format via digitization. Steps indicated by solid lines are implemented, steps indicated by dashed lines are envisioned for future projects.}
    \label{fig:ESCAPE35_overview}
\end{figure}

Figure~\ref{fig:ESCAPE35_overview} summarizes our proposed methodology for P\&ID autocorrection. First, a smart P\&ID is converted into a DEXPI P\&ID according to the DEXPI standard. Alternatively, paper-based P\&IDs can be digitized. Then, a P\&ID graph is generated from the DEXPI P\&ID using our pyDEXPI package~\citep{Goldstein2025_PydexpiPythonFramework}. We utilize the P\&ID graph to facilitate the application of rule graphs which implement various engineering rules for error detection and correction. Next, the rule-based autocorrection yields a corrected P\&ID graph. Ultimately, the corrected P\&ID graph is presented in a P\&ID assistant and transformed back into a DEXPI P\&ID. This manuscript focuses on the part from a DEXPI P\&ID to a corrected P\&ID graph.

\subsection{DEXPI P\&ID to P\&ID graph}

The transformation from the DEXPI P\&ID to the P\&ID graph is achieved through our pyDEXPI package~\citep{Goldstein2025_PydexpiPythonFramework}. The pyDEXPI package is a Python implementation of the DEXPI data model, containing data classes that have attribute relationships to encode composition or reference. Furthermore, the pyDEXPI package includes functions for importing and exporting DEXPI data, such as converting a DEXPI P\&ID to a P\&ID graph. In the P\&ID graph, nodes represent P\&ID components and directed edges represent pipes and signals. The direction of the edge is determined by the material flow or signal direction. To keep the P\&ID graph concise, we only include key P\&ID components. Specifically, we include the following components: (i) plant items with a tag, (ii) piping components, such as valves, (iii) actuators, and (iv) process instrumentation. P\&ID components not considered mainly store connectivity information which we represent as directed edges.

\subsection{Applying rule graphs to a P\&ID graph}

\begin{figure}
    \centering
    \includegraphics[width=\linewidth]{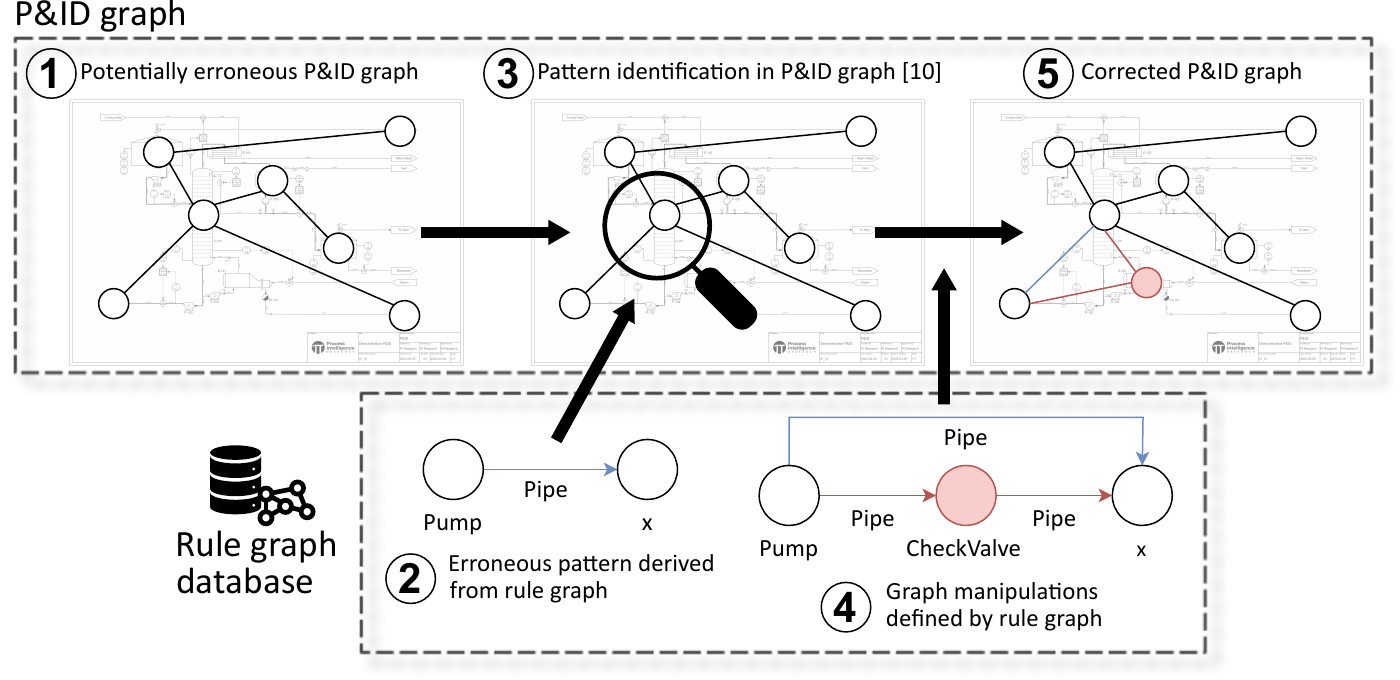}
    \caption{Rule graph execution for Rule 19 “Install a check valve in the pump's discharge line”. According to P\&ID markups, we denote added edges or nodes in red and deletions in blue. Note that for missing components, we search for both the erroneous and the corrected P\&ID pattern (i.e., erroneous P\&ID pattern with manipulations applied) in the P\&ID graph. This is omitted in the figure for brevity.}
    \label{fig:ESCAPE35_graph_rule_check_valve}
\end{figure}

The rule-based autocorrection method searches for erroneous patterns in P\&IDs. If an erroneous pattern is detected, a set of corrections is performed on the pattern. Erroneous patterns and the corresponding set of corrections are collected together in so-called rule graphs. At execution, all rules graphs from a rule graph database are applied sequentially as described in~\Cref{fig:ESCAPE35_graph_rule_check_valve}. 
First, we search for the erroneous graph pattern defined by a rule in the P\&ID graph via subgraph isomorphism. The NetworkX package provides a subgraph isomorphism algorithm based on the VF2 algorithm~\citep{Cordella2004_SubGraphIsomorphism}. VF2 is a recursive algorithm that starts by randomly mapping a node from a pattern to a node in the P\&ID graph with the same DEXPI class. At each step, the algorithm then tries to include pairs of neighboring nodes in the mapping that also have the same DEXPI class while avoiding previously explored mappings. If the complete pattern is mapped to the P\&ID graph, a match is found. There can be multiple matches within a P\&ID graph. We extend the work of \citet{Bayer2019_GraphBasedManipulation} by incorporating conditions based on conditional statements during the subgraph isomorphism. The conditions allow us to go beyond exact attribute matches and incorporate inequality constraints, ranges, and sets. For example, Rule 3 in Table~\ref{tab:ESCAPE35_rules_table} defines one erroneous P\&ID pattern for all line sizes greater or equal to 100 DN (or 4”), instead of defining a separate pattern for every possible line size. Isomorphism conditions are stored as node or edge attributes in the rule graphs. 
Second, we correct the P\&ID graph according to the rule graph for every match that is found. The graph manipulations include nodes and edges that perform either insertion or deletion. Insertion adds nodes or edges to the P\&ID graph, while deletion removes nodes or edges from the P\&ID graph. When visualizing a rule graph, insertion is denoted with red and deletion with blue. 
A special case involves rules that describe missing components. Here, we search for both the erroneous and the corrected P\&ID pattern (i.e., erroneous P\&ID pattern with manipulations applied) in the P\&ID graph. If a match of an erroneous pattern is included within a match with the corresponding corrected pattern, applying the rule is not necessary because the missing component already exists. However, if a match is found that includes only the erroneous but not the corrected pattern, a component is missing and the graph manipulations should be applied.
Figure~\ref{fig:ESCAPE35_graph_rule_check_valve} (step 4) illustrates an exemplary rule graph to insert a check valve in the pump's discharge line (Rule 19 in Table~\ref{tab:ESCAPE35_rules_table}). The rule graph breaks the connection between the pump and the subsequent component “x”. Furthermore, a check valve is added to the graph. Two new connections complete the rule graph: A pipe from the pump to the check valve and a pipe from the check valve to the subsequent component. Here, the “x” node can be mapped to any node in the P\&ID graph. In addition, a pattern node’s DEXPI class can always be mapped to its respective subclasses, i.e., the pump in the rule graph can be mapped to a reciprocating pump in the P\&ID graph.

Besides the graph pattern and manipulations, each rule includes further graph attributes: (i) An ID to identify the rule, (ii) the P\&ID milestone at which the rule is relevant, for instance, issue for review, issue for design, or issue for construction~\citep{Toghraei2019_PipingInstrumentationDiagram}, (iii) a description on how the P\&ID should be revised, (iv) an explanation why the P\&ID should be revised, (v) the level of recommendation, for instance, the rule is a mandatory revision, a suggestion, or only a consideration, (vi) a boolean stating whether the rule describes a missing component, and (vii) the source of the rule.

In total, we define 33 rules. We define the rules based on engineering heuristics, literature~\citep{Woolf2009_ChemicalProcessDynamics}, and existing workflows at Flour©. Here, we describe five exemplary rules in Table~\ref{tab:ESCAPE35_rules_table}.
The output of the methodology developed in this manuscript is a corrected P\&ID graph and the meta data of the applied rules as given in Table~\ref{tab:ESCAPE35_rules_table}.

\begin{table}
    \centering
    \caption{List of rules implemented for the case study.}
    \begin{tabular}{lp{5cm}p{5cm}p{2cm}}
        \toprule
        Rule ID& Rule description & Rule explanation & Implementing the rule is ...  \\
        \midrule
        3	& Do not install a globe valve as a control valve if the pipe diameter is greater or equal to 100 DN (or 4”).	& Large globe valves have higher costs compared to other valve types.	&suggested.\\
        9	&Install a level instrument on a vessel.	&Monitoring the vessel level regularly can prevent accidents caused by overflow.	&mandatory.\\
        10	&Install a strainer in the suction line of a pump.&	The strainer separates solid matter, which can potentially damage the pump, from the fluid.	&suggested.\\
        19	&Install a check valve on a pump's discharge line to avoid backflow.	&Backflow is dangerous to the pump because the pump is designed for one-way flow.	&suggested.\\
        21&	Install block valves and a drain in the suction and discharge of a pump.	&Isolate the pump during maintenance.	&mandatory.\\
        \bottomrule
    \end{tabular}
    \label{tab:ESCAPE35_rules_table}
\end{table}

\section{Results and discussion}

\begin{figure}
    \centering
    \includegraphics[width=0.8\linewidth]{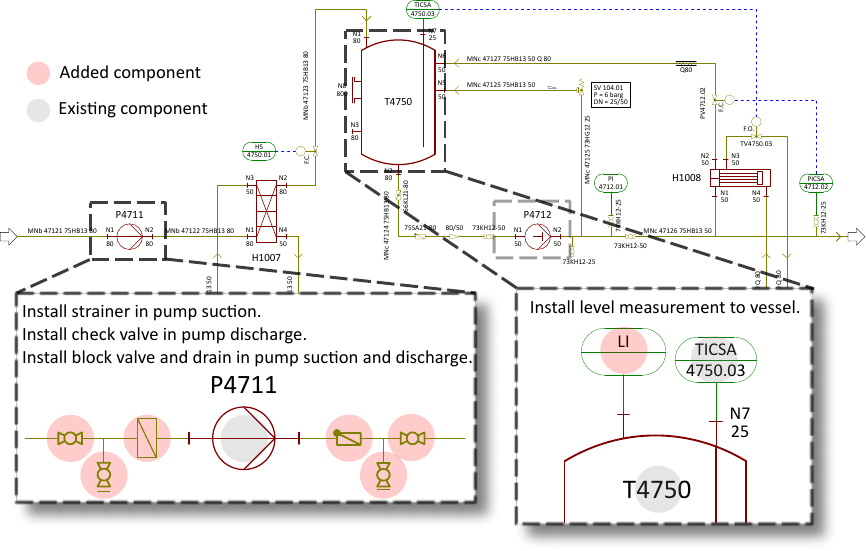}
    \caption{Case study for the rule-based autocorrection using the DEXPI v1.3 reference P\&ID C01 (https://gitlab.com/dexpi/TrainingTestCases). Note that the P\&ID does not reassemble an actual process but acts as a demonstration of the DEXPI data model. The autocorrection method updates pump P4711, pump P4712, and vessel T4750. The same corrections shown for pump P4711 are also applied to pump P4712.}
    \label{fig:ESCAPE35_case_study}
\end{figure}

To evaluate the performance of the autocorrection method, we applied the five rules listed in Table~\ref{tab:ESCAPE35_rules_table} in an illustrative case study by the DEXPI initiative (Figure~\ref{fig:ESCAPE35_case_study}). The P\&ID includes two pumps, two heat exchangers, and a central vessel, with associated instrumentation for pressure and temperature regulation. The process begins with fluid entering through a pump, passing through a heat exchanger for preconditioning, and then entering a vessel for storage or processing. Post-vessel, the fluid is discharged by another pump and is partly recycled to the vessel. The recycle undergoes additional thermal conditioning for temperature control of the vessel. The resulting P\&ID graph consists of 33 nodes and 36 edges. Please note that the case study is not representative for the complexity of a standard P\&ID in the process industry.

\subsection{Detection and correction accuracy}

The autocorrection method demonstrates high accuracy in identifying and correcting errors. The method successfully identified all erroneous substructures from the rule graph database, achieving an accuracy of 100\%. These substructures were corrected using the corresponding rule graphs. For instance, the vessel is lacking pressure and level control. Rule 9 added a level instrument to the vessel for level control. Additionally, the two pumps are not protected and cannot be isolated for maintenance. Rule 21 facilitates maintenance by isolating the pumps with block valves and drains. To protect the pumps, Rule 10 recommended installing a strainer in the suction line, and Rule 19 suggested placing a check valve in the discharge line. For every applied rule, the method provided an explanation for why the revision was necessary. However, the rule graph database did not address all potential errors; for example, the vessel and heat exchangers were not isolated with block valves and drains, indicating the need for additional rules.

\subsection{Computational performance}

The autocorrection method demonstrated computational efficiency. It executed in 0.016 seconds on a laptop equipped with an i7-1185G7 processor (1.8-3.0 GHz) and 16 GB RAM. On average, evaluating a single rule took only 3.2 milliseconds. This performance is significantly faster than manual detection and correction of errors, which can take several minutes to hours, particularly in cluttered P\&IDs. 

\subsection{Discussion}

Our findings indicate that the rule-based autocorrection method provides reliable and efficient support for engineers. Its ability to identify and correct errors with deterministic accuracy makes it suitable for safety-critical applications. The rapid execution time also highlights its potential for real-time applications, saving engineers significant time compared to manual reviews. Moreover, the method’s integration with smart P\&IDs enables it to utilize detailed component attributes, reducing the need for external documentation.
This study emphasizes the importance of applying rules in the correct sequence. For example, applying Rule 21 before Rules 10 and 19 ensures proper placement of strainers and check valves near pumps. Selecting the correct sequence minimizes the risk of overlapping or conflicting corrections.
However, the method’s scalability is constrained by the need for manual rule derivation and maintenance. Each rule requires expertise from trained engineers, and as the rule database expands, tracking and managing interactions between rules becomes increasingly complex. In contrast, ML-based approaches excel at scaling, offering a promising avenue for future research~\citep{Schweidtmann2024_GenerativeArtificialIntelligence}. Exploring hybrid methods that combine rule-based and ML-based approaches could leverage the strengths of both methodologies~\citep{Schweidtmann2021_MachineLearningChemical}.
In conclusion, while the rule-based autocorrection method is not a standalone tool, it serves as a valuable support system for engineers, enhancing the accuracy and efficiency of P\&ID reviews.

\section{Conclusions}

This study presents rule-based autocorrection in DEXPI P\&IDs. First, a DEXPI-formatted P\&ID is converted into a P\&ID graph using pyDEXPI. Subsequently, rule graphs based on chemical engineering knowledge and heuristics are applied to the P\&ID graph to detect and rectify errors. The final outcome is a corrected P\&ID graph. We implement five selected rules and test them in a case study. The results show that the autocorrection method has an accuracy of 100\% due to the rule-based approach. Therefore, the method is reliable and applicable in safety-critical applications. Besides this, the method detects and corrects errors at a significantly faster rate than humans. This research tackles previous limitations by (i) being compatible with the data exchange standard DEXPI through our pyDEXPI module, (ii) proposing rule graphs embedding engineering experience, and (iii) capturing missing components. However, the rule-based method also comes with limitations. Engineers need to extend and maintain the list of rules. In addition, rules may conflict. Therefore, the order in which we apply the rules and their context is important. Future research should investigate hybrid methods combining rule-based and ML-based for autocorrection. In addition, completing the autocorrection pipeline from Figure~\ref{fig:ESCAPE35_overview} by implementing the conversion of P\&ID graphs to the DEXPI data model would enable the full integration into existing engineering workflows.

\section*{Acknowledgments}

We gratefully acknowledge the support provided by Linde GmbH, Linde Engineering and Siemens Aktiengesellschaft (DPG), as well as Flour© and Top Sector Alliance for Knowledge and Innovation (TKI) by the Dutch Ministry of Economic Affairs and Climate (CHEMIE.PGT.2023.033) (LSB).

\bibliography{references}

\end{document}